\documentclass[12pt]{iopart}
\usepackage{iopams}  
\usepackage{graphicx}
\usepackage{setstack}
\usepackage{color}
\usepackage{relsize}
\usepackage{hyperref}

\DeclareMathAlphabet{\mathitbf}{T1}{cmr}{bx}{it} 

\begin{document}
  
\title[Numerical study of the overlap Lee-Yang singularities in the EA model]{Numerical Study of the Overlap 
Lee-Yang Singularities in the
  Three-Dimensional Edwards-Anderson Model}

\author{R. A. Ba\~nos$^{1,2}$, J. M. Gil-Narvion$^{2}$,
        J.~Monforte-Garcia$^{1,2}$, J.~J.~Ruiz-Lorenzo$^{3,2}$
        and D.~Yllanes$^{4.2}$}
\address{$^1$ Departamento de F\'{\i}sica Te\'orica, Universidad de Zaragoza, 50009 Zaragoza, Spain.}
\address{$^2$ Instituto de Biocomputaci\'on y F{\'{\i}}sica de Sistemas Complejos (BIFI), 50018 Zaragoza, Spain.}
\address{$^3$ Departamento de F\'{\i}sica, Universidad de Extremadura, 06071 Badajoz, Spain.}
\address{$^4$ Dipartimento di Fisica, La Sapienza Universit\`a di Roma, P.le A. Moro 2, 00185 Roma, Italy.}

\date{\today}

\begin{abstract}

We have characterized numerically, using the Janus computer, the Lee-Yang
complex singularities related to the overlap in the $3D$ Ising spin glass
with binary couplings in a wide range of temperatures (both in the critical
and in the spin-glass phase). Studying the behavior of the zeros at the
critical point, we have obtained an accurate measurement of the anomalous
dimension in very good agreement with the values quoted in the literature.
In addition, by studying the density of the zeros we have been able to
characterize the phase transition and to investigate the Edwards-Anderson order
parameter in the spin-glass phase, finding agreement with the values
obtained using more conventional techniques.

\end{abstract}


\pacs{75.10.Nr,71.55.Jv,05.70.Fh}

\maketitle

\section{Introduction}


In two seminal papers, T. D. Lee and C. N. Yang \cite{yang:52,lee:52}
introduced a new tool to understand the origin of a phase transition by
studying the complex singularities of the free energy, or, equivalently, the
zeros of the partition function. In particular, they showed that all the zeros
are located on the unit circumference on the complex activity plane (taking as
variable $z=e^{-2 h}$, where $h$ is the external magnetic field).  They also
proved that if the zeros condense onto the real axis when $V \rightarrow
\infty$ a phase transition takes place. Finally, they related, in the
low-temperature phase, the density of zeros with the discontinuity in the
order parameter (remember that the Ising model experiences a first-order phase
transition when $h$ changes at a fixed temperature below the critical
one). This approach was subsequently extended to the temperature zeros by
Fisher~\cite{fisher:65,one:68,falcioni:82,marinari:84}.


We have Lee-Yang like theorems only for a limited class of non-disordered
systems (such as Ising models). However, it is possible to develop a
scaling theory by assuming that asymptotically the complex singularities
(wherever they lie) touch the real axis (thus generating the phase
transition).  Hence, despite the lack of formal theorems it is still
possible to apply Lee and Yang's main results to a wide class of systems
(e.g., Potts models~\cite{kenna:06}). In this class of systems the zeros do not live on a
circle as stated by the Lee-Yang theorem, but they still control the
critical properties of the model. We will only assume this last fact
irrespectively of the form of the locus of the zeros in the complex plane~\cite{kenna:06}.


In Refs.~\cite{ruiz-lorenzo:97,kenna:08,gordillo-guerrero:09} the analysis
of complex singularities was applied successfully to diluted systems (in
particular diluted Ising models in two and four dimensions). The key point
for the applicability of the standard results, well tested in
non-disordered systems, is to compute the complex singularities
individually for each disorder realization (called sample) and then use the
mean of the individual zeros (sample zeros) in order to test the scaling
properties of the zeros and  to study the properties of the integrated
density of zeros. In this work we will also introduce the analysis of the
median.


Nowadays  we are interested in gaining a deeper understanding (from the
point of view of the complex singularities) of the properties of an
interesting frustrated and disordered system: the three-dimensional Ising spin
glass. The magnetization, while very interesting in off-equilibrium dynamics
and in experiments, plays no role in the critical behavior and in the
understanding of the low-temperature properties in a finite-dimensional spin
glass. The observable that controls this spin-glass phase is the overlap.
Hence, in this work we have focused on the numerical study of the complex
singularities linked with the overlap in order to study the phase transition
and the properties of the spin-glass phase.

In the past, Lee-Yang and Fisher zeros were obtained in spin glasses by
means of the numerical evaluation of the partition function on small
lattices~\cite{ozeki:88,damgaard:95,bhanot:93,saul:93}.  This methodology
was also applied to models defined on Bethe lattices~\cite{matsuda:10}.
Finally, some calculations were performed with the help of
replicas~\cite{takahashi:11}.

More recently, the complex singularities linked with the external magnetic
field were studied for the two and three-dimensional Ising spin glass model in
the interesting reference~\cite{matsuda:08}, which focuses on the Griffiths
singularity and computes all the zeros for small lattices.

In particular we are interested in characterizing the scaling of the
individual zeros at the critical point (which will allow us to compute the
anomalous dimension exponent) and checking the scaling in the spin-glass
region. In addition, we want to study the properties of the density of zeros
in the critical and spin-glass region: the behavior of this observable will
clearly signal the phase transition. Finally, we will show
how this density of zeros can be used to compute the Edwards-Anderson order
parameter. However, the spin-glass susceptibility presents strong scaling
corrections (even on an $L=32$ lattice and $\beta=1.4$), which induce
strong corrections on the density of zeros allowing us (from the numerical
point of view) only to test our density of zeros against the values of
$q_\mathrm{EA}$ found in the literature, rather than attempting a direct numerical computation
of the order parameter. We want to stress that in cases in which the spin
glass susceptibility reaches the asymptotic value, the method we propose will
be able to provide directly the order parameter ($q_\mathrm{EA}$) giving
an additional method to those used
nowadays~\cite{herisson:02,janus:09b,janus:10,janus:10b}.


Let us finally mention that we have obtained the data presented in this work
from the analysis of the configurations produced in parallel tempering
runs~\cite{janus:10,janus:10b} using the Janus
computer~\cite{janus:08,janus:09,janus:12b}.

\section{Model and  observables}\label{sec:MODEL}

We have studied the three-dimensional Edwards-Anderson model with
dynamical variables $\sigma_i$.  These variables are Ising spins and are
placed on the nodes of a cubic lattice of linear dimension $L$ and volume
$V=L^3$.  The Hamiltonian of the system is
\begin{equation}
 \displaystyle\mathcal{H}_0=-\sum_{\langle i,j \rangle}
J_{ij}\sigma_i\sigma_j,
\end{equation}
where $\langle ij \rangle$ indicates that the sum is over the nearest
neighbouring sites.  The couplings $J_{ij}$ are random quenched constants
with bimodal probability distribution, that is, $J=\pm1$ with $50\%$
probability. Every realization of the couplings is called a sample.  Due to
the fact that we have a random Hamiltonian, we have to deal with a double
average: first the thermal average, which  we will denote by $\langle ( \cdots )
\rangle$, and then the average over the samples, which we will denote by
$\overline{(\cdots)}$.

We have simulated several real replicas of the system,
so we can define the local overlap
\begin{equation}
 q_i=\sigma_i^{(1)}\sigma_i^{(2)}
\end{equation}
where $\sigma_x^{(1)}$ belongs to the first replica and $\sigma_x^{(2)}$
belongs to the second one.  The spin overlap is defined from this local
overlap as
\begin{equation}
 \displaystyle Q=\sum_i q_i,
\end{equation}
where the sum runs over the whole volume ($V$).  In addition, we define $q\equiv Q/V$. These observables allow us to
define some new quantities, for example the non-connected spin-glass
susceptibility
\begin{equation}
\chi_{\mathrm{SG}}=\frac{1}{V}\overline{\langle Q^2\rangle}.
\end{equation}

Let us now rewrite the Hamiltonian adding a new perturbation $\epsilon Q$
and including the two replicas explicitly,
 \begin{eqnarray}
 \displaystyle\mathcal{H}_\epsilon&=&\mathcal{H}(\sigma_1)+\mathcal{H}(\sigma_2)+\epsilon Q \nonumber\\
&=&-\sum_{\langle i,j \rangle} J_{ij}\left(\sigma_i^{(1)}\sigma_j^{(1)}+\sigma_i^{(2)}\sigma_j^{(2)}\right) 
+\epsilon\sum_i \sigma_i^{(1)}\sigma_i^{(2)}.
\label{eq:epsilon}
\end{eqnarray}
This Hamiltonian looks like that of the Ising model in a magnetic field
\begin{equation}
 \displaystyle \mathcal{H}_h=\mathcal{H}_0+h\mathcal{M}.
\end{equation}
  We can write the partition function, whose zeros we want to study, as
\begin{eqnarray}
 \displaystyle Z&=&\sum_{[\sigma^{(1)}\sigma^{(2)}]} e^{-\beta \mathcal{H}_0+\beta i \epsilon Q}\\
&=&\sum_{[\sigma^{(1)}\sigma^{(2)}]} \left( \cos(\beta\epsilon Q) e^{-\beta
   \mathcal{H}_0} 
+ i \sin(\beta\epsilon Q) e^{-\beta \mathcal{H}_0} \right). \nonumber
\end{eqnarray}
Let $Z_0$ the partition function of the non-perturbed system, so
\begin{equation}
\displaystyle Z=Z_0 \lbrace \langle \cos(\beta\epsilon Q) \rangle + i \langle \sin(\beta\epsilon Q)\rbrace
\end{equation}
and we have to find the zeros of the function $\langle \cos(\beta\epsilon
Q) \rangle$ since in absence of a magnetic field $\langle
\sin(\beta\epsilon Q) \rangle$ is zero.  The algorithm to find them is
quite easy: we start from the list of individual measurements of $Q$ for
each sample (see Section~\ref{sec:simulations}) and evaluate the average
$\langle \cos(\beta \epsilon Q)\rangle$, increasing $\epsilon$ in small
steps $\Delta \epsilon$.  When the function changes signs from one step to
the next, we have found a zero in this interval.  Obviously, the smaller
$\Delta\epsilon$ the better the precision of the zero that we have found,
but also the slower the analysis, so we have to be careful with the error
estimates.  We have analyzed the first four zeros of this function.

\section{Finite-Size Scaling}\label{sec:FSS}

One can obtain the expected behavior of the LY zeros by means of (see for
example Ref. ~\cite{kenna:06}) 
\begin{equation}
\epsilon \simeq \frac{1}{\sqrt{\chi_\mathrm{SG} V}}\,
\label{eq:scaling_e}
\end{equation}
therefore, the finite-size dependence, at the critical point, of the
Yang-Lee zeros can be expressed as:
\begin{equation}\label{eq:scaling-Tc}
 \epsilon_j (L) \sim L^{-x_1} \,,
\label{eq:scaling}
\end{equation}
where 
\begin{equation}\label{eq:x1}
x_1=(D+2-\eta)/2\,, 
\label{scaling_x1}
\end{equation}
and  $D$ is the dimensionality of the system, being $D=3$ in this work.
If corrections to scaling are taken into account, the above relation becomes
\begin{equation}\label{eq:scaling-Tc-corr}
 \epsilon_j (L) \sim L^{-x_1}\left(1+{\cal O} (L^{-x_2})\right) \,,
\label{eq:corrections_to_scaling}
\end{equation}
where $x_2$ is the leading correction-to-scaling exponent, $x_2=\omega$. 

In the broken symmetry phase, where the non-linear susceptibility diverges
as the volume of the system, we expect the following behavior:
\begin{equation}\label{eq:scaling-lowT}
 \epsilon_j (L) \sim \frac{1}{V} \,,
\label{eq:scaling_below}
\end{equation}
We can take scaling corrections into account, as in the critical point, and
\begin{equation}\label{eq:scaling-lowT-corr}
 \epsilon_j (L) \sim V^{-1}\left(1+{\cal O} (L^{-x_3})\right) \,.
\label{eq:corrections_to_scaling_below}
\end{equation}
where $x_3$ is the leading correction-to-scaling exponent in the broken
phase.\footnote{Both droplet and RSB predict algebraic decays for the connected
  correlation functions in the spin glass phase (the spin glass phase is
  critical in both models). In the droplet model the exponent of the decay is
  $y$ (sometimes denoted as $\theta$), so one can show that $x_3=y$. In RSB
  depending of the value of $q$ we have different decays (of the $q$-constrained
  correlation functions), denoting the decay exponents as $\theta(q)$. So the
  leading correction exponent can be shown to be the smallest of the different
  $\theta(q)$. See Refs.~\cite{janus:10,janus:10b} for a detailed discussion on $\theta(q)$.}.

In order to discuss the density of zeros we need to describe some known
properties of the Hamiltonian defined in Eq. (\ref{eq:epsilon}). This
Hamiltonian was introduced in the past~\cite{parisi:89,franz:92}.  In
particular it experiences a first-order phase transition in $\epsilon$,
below the critical temperature of the uncoupled model. Hence, the overlap
is discontinuous:
\begin{equation}
\lim_{\epsilon \to 0^{\pm}} \overline{\langle q \rangle}(\epsilon)=\pm
q_\mathrm{EA}\,,
\end{equation}
being the discontinuity at the transition just $2 q_\mathrm{EA}$.

We can also introduce the density of zeros
\begin{equation}
\mu_\epsilon(\epsilon)=\sum_j \delta(\epsilon-\epsilon_j(L))
\label{eq:density}
\end{equation}
and its integrated version 
\begin{equation}
G(\epsilon)=\int_0^\epsilon dx \mu_\epsilon(x) 
\end{equation}
which takes the following value computed for a given zero:
\begin{equation}
G(\epsilon_j(L))=\frac{2 j -1}{2 V} \,,
\end{equation}
where $j$ labels the $j$-th zero ($j=1, 2,\ldots$). In order to deal with
the discontinuous behavior of $G(\epsilon)$ at the zeros, we follow the
recipe of references~\cite{janke:01,janke:04} and use the mean between two
consecutive plateau values ($j-1$ and $j$). Anyhow, the asymptotic value
of the integrated density computed in the $j$-th zero is $j/V$. We will
discuss this point again in subsection \ref{sub:density}.

This integrated density
is very useful to characterize a phase transition. In general it behaves as
\begin{equation}
G(\epsilon)=a_1 \epsilon^{a_2}+a_3
\end{equation}
and we can extract a great amount of physical information from these three
numbers ($a_1, a_2$ and $a_3$):
\begin{itemize}
\item In the symmetric phase $a_3<0$. In a broken phase $a_3>0$.
\item In the onset of a first-order phase transition, varying $\epsilon$ as it is
  our case: $a_2=1$ and $a_3=0$. In addition we can extract the order
  parameter of the broken phase: $q_\mathrm{EA}=\pi a_1/\beta$.\footnote{In
    Lee and Yang's paper, the starting point is the Hamiltonian $\beta {\cal H}_h=\beta
    {\cal H}_{h=0} + h M$, where $M$ is the total magnetization of the system. In
    terms of the fugacity $z=e^{-2h}$, they obtained the following result
    (valid below the critical temperature) for the density of zeros (in the
    fugacity variable that we will denote as $\mu_z(z)$):
    $\mu_z(0)=m_\mathrm{sp}/(2 \pi)$, where $m_\mathrm{sp}$ is the spontaneous
    magnetization below the critical temperature. In order to transfer this
    result to our notation we remark that our ``magnetic field'' is $\beta
    \epsilon$, $q_\mathrm{EA}$ plays the role of $m_\mathrm{sp}$ and we need to use the
    standard law of the transformation of the probability densities
    ($\mu_z(z)=\mu_\epsilon(\epsilon) |d\epsilon/dz|$, where $z=\exp(-2 \beta \epsilon)$), obtaining:
\begin{equation}
\mu_\epsilon(0)=\frac{q_\mathrm{EA} \beta}{\pi} \,.
\end{equation} Notice that near $\epsilon=0$ we can identify $a_1$ with $\mu_\epsilon(0)$.
}

\item At the critical point, $a_3=0$ and $a_2$ is related with the anomalous
  dimension $\eta$ by means:
\begin{equation}
a_2=\frac{2 D}{D+2-\eta}\,.
\label{scaling_a2}
\end{equation}

\end{itemize}

\section{Simulation details}\label{sec:simulations}

We have run simulations for several lattice
sizes on the Janus supercomputer~\cite{janus:08,janus:09,janus:12b} (for
$L=16, 24, \mbox{ and } 32$) and on conventional computers (for $L=8 \mbox{
and } 12$). These simulations were originally reported in~\cite{janus:10},
which gives full details on the chosen parameters and the thermalization
protocol. In this section we give only a brief summary.

We have used the parallel tempering algorithm
\cite{hukushima:96,marinari:98b}, choosing  the temperatures to maintain
an acceptance around $20\%$ in parallel tempering updates.  Besides, since
Janus needs far more time to do a parallel tempering update than a
heat-bath one, we have chosen to do one parallel tempering update every
$10$ heat-bath ones.  In table \ref{tab:summary-simulations} one can find a
summary of the simulations parameters. In order to choose
the simulation length, we have assessed thermalization 
on a sample-by-sample basis, using the temperature random walk
technique~\cite{fernandez:09b,janus:10}
(table~\ref{tab:summary-simulations} gives the average number of lattice
updates for each $L$).

In general, each of the single processors (FPGAs) of Janus takes care of
the simulation of one replica of the system. However, some samples have
such a slow dynamics that even with this algorithm the simulation would be
too long (more than six months of continuous running time), so we would
need to accelerate it. For these few cases we have created a special
low-level code that is in charge of the parallel tempering in the control
FPGA of a board of Janus.  This allows us to spread the simulation over
several processors running only a subset of temperatures in each FPGA, thus
accelerating the simulation by increasing the parallelism.

\begin{table} [!ht]
\caption{Summary of the simulations.  $N_\mathrm{T}$ is the number of
simulated temperatures (evenly spaced between $T_\mathrm{min}$ and
$T_\mathrm{max}$); $N_\mathrm{mes}$ is the number of Monte Carlo steps
(updates of the whole lattice) between measurements; 
$N_\mathrm{HB}^\mathrm{med}$ is the average simulation time (since we use
the random-walk technique the simulation   time depends on the sample);
$N_\mathrm{sam}$ is the number of simulated samples.  We have simulated
four real replicas for each sample. Finally, $L=8$ and $L=12$ have been
simulated on PCs and $L=16$, $L=24$ and $L=32$ on Janus.\label{tab:summary-simulations}}
 \begin{tabular*}{\columnwidth}{@{\extracolsep{\fill}}ccccccc}
\br
L&$T_\mathrm{min}$&$T_\mathrm{max}$&$N_\mathrm{T}$&$N_\mathrm{mes}$&$N_\mathrm{HB}^\mathrm{med}$&$N_\mathrm{sam}$\\
\mr
$8$&$ 0.150 $&$1.575$&$ 10$ & $10^3$& $ 7.82 \times 10^6 $ & $4000$ \\ 
$12$&$0.414$&$1.575$&$12$&$5 \times 10^3$&$3.13 \times 10^7$ & $4000$ \\
$16$&$0.479$&$1.575$&$16$&$10^5$&$9.71 \times 10^8$ & $4000$ \\
$24$&$0.625$&$1.600$&$28$&$10^5$&$4.02 \times 10^9$ & $4000$ \\
$32$&$0.703 $&$ 1.549 $&$ 34 $&$2 \times 10^5$&$ 1.90 \times 10^{10} $ & $1000$ \\
\br
 \end{tabular*}
\end{table}

\subsection{Data for the computation of the zeros}
We have saved on disk every individual measurement of the overlap.  Since we
have simulated four real replicas of the system, for each sample we have a
total of $6N_\mathrm{HB}/N_\mathrm{mes}$ values of $Q$.  Given the variable
$N_\mathrm{HB}$, this ranges from $1.2\times10^5$ to $2\times10^7$ measurements
for our largest lattices, so we have a very good precision for computing the
zeros of the partition function. We have discarded the first half of the
measurements for equilibration.

We want to study the behavior of the system in the critical temperature and
in the low-temperature phase of the system, analyzing the scaling of the
zeros.  Therefore, we need to compute the zeros of different system linear
sizes, $L$, but at the same temperature. Since we have not simulated the
same temperatures for every lattice size, we have interpolated, using cubic
splines, in order to estimate the zero at each of the chosen scaling
temperatures.

\section{Results}\label{sec:RESULTS}

In this section we will study the behavior of the zeros as a function of the
lattice size, both in the critical and in the spin-glass phase. Finally,
we will compute the density of zeros and extract the $\eta$ exponent from the
analysis at the critical temperature and the Edwards-Anderson order parameter
from the scaling in the low-temperature phase.

\subsection{Scaling at the Critical Point}\label{sub:scaling_tc}

We first consider the scaling at the critical point and use it to determine the
anomalous dimension, as in~(\ref{eq:scaling-Tc}).  Our simulations were
optimized to investigate the low-temperature phase, for large system sizes,
rather than to obtain the critical parameters. Therefore, we take the value of
$\beta_\mathrm{c}=0.902(8)$ from~\cite{hasenbusch:08b}, which features many
more samples and small sizes to control scaling corrections but does not reach
the low-temperature phase, and will also use this reference to check our value
of $\eta$.\footnote{If we combine the critical exponents
of~\cite{hasenbusch:08b} with the Janus simulations studied herein, we obtain a
compatible value of $\beta_\mathrm{c}=0.905(7)$~\cite{billoire:11}.}

Let us first consider a fit of the individual zeros,
leaving aside corrections to 
scaling, i.e., following~(\ref{eq:scaling-Tc}). For the $j$-th zero, we fit to
\begin{equation}\label{eq:fit-Tc}
\epsilon_j(L) = A_j L^{-x_1}.
\end{equation}
We report the results of these fits in table~\ref{table:scaling_zeros_tc}. We
see that the first and second zeros follow~(\ref{eq:fit-Tc}) very well for
$L\geq8$, but for $j>2$ we need to restrict the fit to $L\geq12$. However,
there is an inconsistency in the results: the value of $x_1$ should be the same
for all zeros, but we see that it increases with $j$. Moreover, at least for
the larger $j$, $x_1$ is incompatible with the expected value of $x_1=2.688(5)$,
(taking $\eta=-0.375(10)$ from~\cite{hasenbusch:08b}) 
This hints that corrections to scaling should be taken into account, 
as in~(\ref{eq:scaling-Tc-corr}).  
\begin{table}[!t]
\centering
\caption{Fits of the zeros to $\epsilon_j(L) = A_j L^{-x_1}$, for $L\geq L_\mathrm{min}$. 
As we can see, with $L_\mathrm{min}=8$ the $\chi^2$ per degree of freedom is acceptable only
for $j=1,2$, but with $L_\mathrm{min}=12$ all the zeros have a reasonable fit.
However, the value of $x_1$ grows with $j$, an indication that we have to consider
corrections to scaling (see text). 
\label{table:scaling_zeros_tc}}
\begin{tabular*}{\columnwidth}{@{\extracolsep{\fill}}ccccc}
\br
$j$ & $L_\mathrm{min}$ &  $\beta$ & $x_1$ &  $\chi^2 /\mathrm{d.o.f.}$ \\
\mr
1 & 8  & 0.902  & 2.703(12) & 1.78/3\\
2 & 8  & 0.902  & 2.712(6)  & 3.23/3\\
3 & 8  & 0.902  & 2.718(5)  & 8.12/3\\
4 & 8  & 0.902  & 2.725(5)  & 15.1/3\\ 
\mr
1 & 12  & 0.902  & 2.695(14) & 1.27/2\\
2 & 12  & 0.902  & 2.715(8)  & 2.95/2\\
3 & 12  & 0.902  & 2.731(7)  & 2.19/2\\
4 & 12  & 0.902  & 2.745(7)  & 2.49/2 \\
\br
\end{tabular*}
\end{table}
\begin{figure}[!ht]
\centering
\includegraphics[height=0.7\textwidth,angle=0]{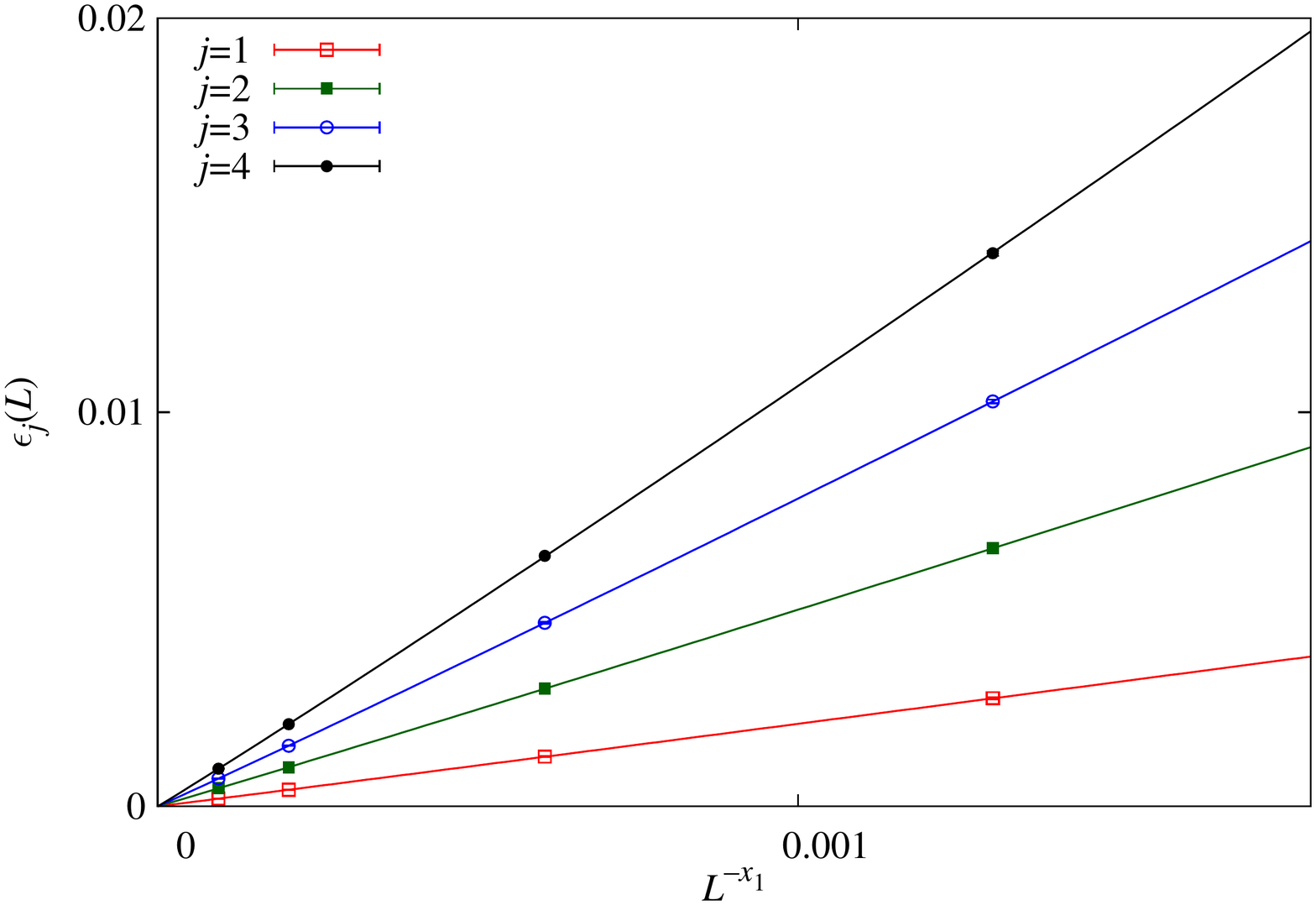}
\caption{The four first zeros at $\beta=\beta_c$. In order to appreciate
the scaling better, we show only the data for $L\geq12$ and 
compare to equation~(\ref{eq:scaling-Tc-corr}), 
fixing $x_2=\omega=1.0(1)$ from~\cite{hasenbusch:08b} and performing a global fit for a 
common value of $x_1$ (see text). We obtain  
$x_1 = 2.67(6)[1]$, with a chi-square per degree of freedom of $\chi^2/\mathrm{d.o.f.} = 5.88/7$.
\label{fig:scaling_tc}}
\end{figure}

In order to do so, we consider all values of $j$ at the same time and perform
a global fit, enforcing data from different zeros to share the same $x_1$ and
$x_2$. As points coming from a given $L$ are correlated, the full covariance
matrix has to be considered.  We label our set
of points $\{\epsilon_j(L_a)\}$ by their $L$ and their $j$: we have data for
$\mathcal L=5$ different values of $L$ ($L_1=8$, $L_2=12$, $L_3=16$, $L_4=24$,
$L_5=32$) and for $j=1,2,3,4$. The appropriate goodness-of-fit estimator is,
therefore, \begin{equation} \hspace{-2cm} \chi^2 = \sum_{i,j=1}^4
\sum_{a,b=1}^{\mathcal L} [\epsilon_i(L_a)- A_i L_a^{-x_1}(1+B_iL_a^{-x_2})]
\sigma_{(ia)(jb)}^{-1} [\epsilon_j(L_b)- A_j L_b^{-x_1}(1+B_jL_b^{-x_2})],
\end{equation} where $\sigma_{(ia)(jb)}$ is the covariance matrix of the set of
zeros (which is block diagonal, since data for different $L$ are uncorrelated).

Unfortunately, we do not have enough data to determine $x_2$ and $x_1$ at the
same time (the resulting error in $\omega$ would be greater than $100\%$).
Instead, we take $x_2=\omega=1.0(1)$ from~\cite{hasenbusch:08b} and fit only
for $x_1$ and the amplitudes. The resulting fit for $L\geq12$, shown in
figure~\ref{fig:scaling_tc}, gives \begin{equation} x_1 = 2.67(6)[1],\qquad
\chi^2/\mathrm{d.o.f.} = 5.88/7, \end{equation} where the error in square
brackets accounts for the uncertainty in $\omega$.  Our determination of $x_1$
is now compatible with the expected value of $x_1=2.688(5)$.  Therefore, the
scaling of the zeros is consistent at the critical point.

\subsection{Scaling in the low-temperature phase}
\label{sec:scaling_low}
Now we consider the scaling of $\epsilon_j(L)$ in the low-temperature phase.
This time, we expect, from~(\ref{eq:scaling-lowT}),
\begin{equation}\label{eq:scaling_lowT2}
\epsilon_j(L) \simeq A L^{-x_1},
\end{equation}
with $x_1=D$.
\begin{figure}[tb]
\centering
\includegraphics[height=0.7\textwidth,angle=270]{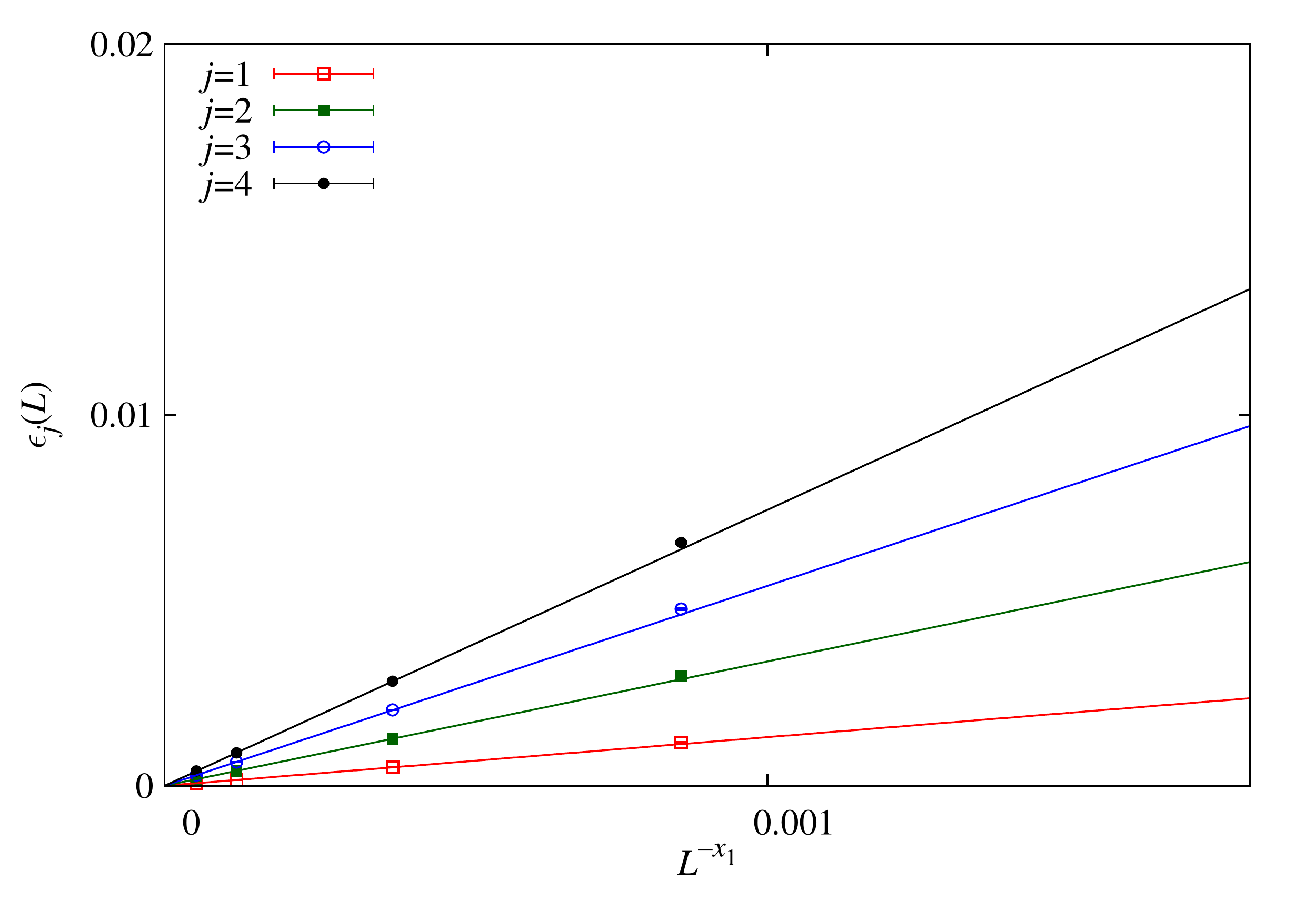}
\caption{Scaling of the zeros at $\beta=1.4$, with a best fit to (\ref{eq:scaling_lowT2}) for $L\geq16$.
We obtain $x_1=2.842(11)$, with $\chi^2/\mathrm{d.o.f.}=7.34/7$.}
\label{fig:scaling_b1.4}

\centering
\includegraphics[height=.7\textwidth,angle=270]{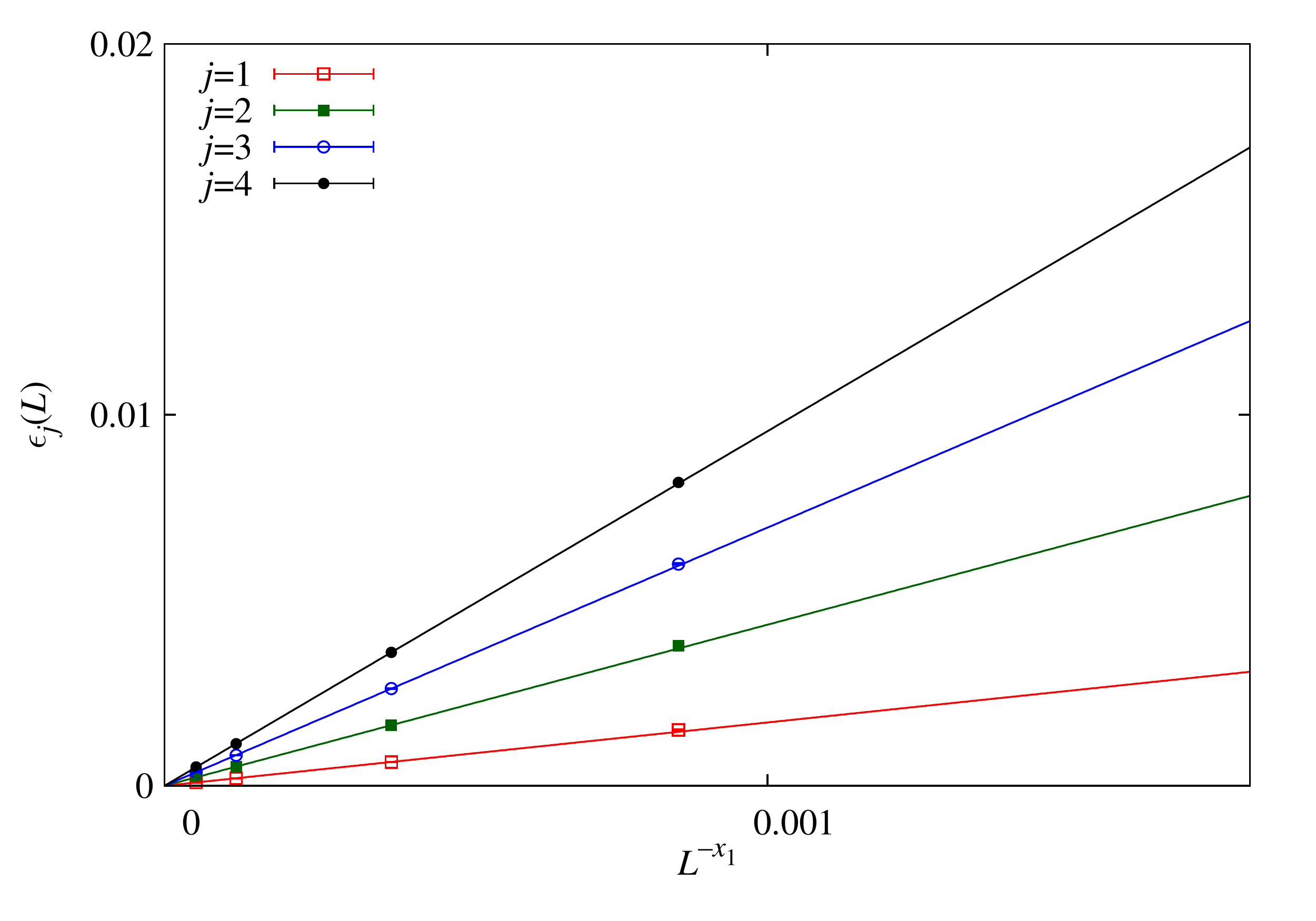}
\caption{Scaling of the zeros at $\beta=1.2$, with a best fit to (\ref{eq:scaling_lowT2}) for $L\geq16$.
We obtain $x_1=2.844(10)$, with $\chi^2/\mathrm{d.o.f.}=2.89/7$.}
\label{fig:scaling_b1.2}
\end{figure}

We have fitted the data for $\beta=1.2$ (Figure~\ref{fig:scaling_b1.2}) and
$\beta=1.4$ (Figure~\ref{fig:scaling_b1.4})
to~(\ref{eq:scaling_lowT2}).\footnote{The crossover length $L_c$ which marks
  the change between the criticality induced by the critical point at $T_c$
  ($L<L_c$) and that of the spin glass phase ($L>L_c$) has been computed for
  different values of $\beta$ in reference~\cite{janus:10}. In particular, we
  know that $L_c(\beta=1.2)\simeq 6$ and $L_c(\beta_c=1.4))\simeq 2.5$. Hence
  all the data presented in this section belong to temperatures which lie deep
  into the spin glass phase. In other words, we can only see the critical
  effects induced by the spin-glass phase itself, which is critical, not those
  of the critical point at $T_c$.}  The results, in
Table~\ref{tab:scaling-lowT}, show a value of $x_1$ incompatible with
$x_1=D=3$. We have also included corrections to scaling, using both $\omega=1$
(Goldstone-like correction)~\cite{gordillo:12} and $\omega=3$ (Ising ordered
correction)~\cite{gordillo:12}, $\omega=y=0.255$
(droplet)~\cite{carter:02,boettcher:04,boettcher:05}, $\omega=\theta(0)=0.39$
(replicon and also $1/\hat{\nu}$ which controls the scaling correction of
$q_\mathrm{EA}(L)$ \cite{janus:10b}), $\omega=0.79=2 \theta(0)$ (twice the
replicon \cite{janus:10b}) and
$\omega=0.65=\theta(q_\mathrm{EA})$~\cite{janus:10b} but in neither case is
the asymptotic $x_1=D$ behavior recovered (see
Table~\ref{tab:scaling-lowT}). In addition, we have forced the fits with
$x_1=3$ and leaving free $\omega$ and the statistical quality of the fits was bad.
\begin{table}[tb]
\centering
\caption{Scaling of the zeros in the low-temperature phase.  For the two
  considered temperatures ($\beta=1.2,1.4$) we first show a fit without
  corrections to scaling for $L\geq16$, that is $\epsilon_j(L) \simeq A_j
  L^{-x_1}$. As explained in Section~\ref{sub:scaling_tc}, this is a global
  fit for the four zeros, considering their full covariance matrix. We then
  consider the same fit with corrections to scaling, trying different values
  for $\omega$ (see the text for more details). In all cases $x_1$ is smaller
  than the expected value $x_1=D=3$.
\label{tab:scaling-lowT}}
\begin{tabular*}{\columnwidth}{@{\extracolsep{\fill}}ccccc}
\br
$L_\mathrm{min}$ &  $\beta$ & $\omega$& $x_1$ &  $\chi^2 /\mathrm{d.o.f.}$ \\
\mr
16 & 1.4 & - & 2.842(11) & 7.34/7\\
12 & 1.4 & 1 & 2.57(12) & 3.79/7\\
12 & 1.4 & 3 & 2.79(2)  & 4.18/7\\
12 & 1.4 & 0.255 & 2.75(10) & 17.6/7\\
12 & 1.4 & 0.39  & 2.67(9) & 14.7/7\\
12 & 1.4 & 0.65  & 2.55(11) & 7.90/7\\ 
12 & 1.4 & 0.79  & 2.48(13) & 5.03/7\\
\mr
16 & 1.2 & - & 2.844(10) & 2.89/7 \\
12 & 1.2 & 1 & 2.82(5)   & 10.5/7\\
12 & 1.2 & 3 & 2.84(2)   & 6.90/7\\
12 & 1.2 & 0.255 & 2.80(10) & 12.3/7\\
12 & 1.2 & 0.39  & 2.80(12)  & 11.9/7\\
12 & 1.2 & 0.65  & 2.81(10)   & 11.3/7\\
12 & 1.2 & 0.78  & 2.81(8)   & 11.0/7\\ 
\br
\end{tabular*}
\end{table}

\begin{figure}[!ht]
\centering
\includegraphics[width=.9\textwidth, angle=0]{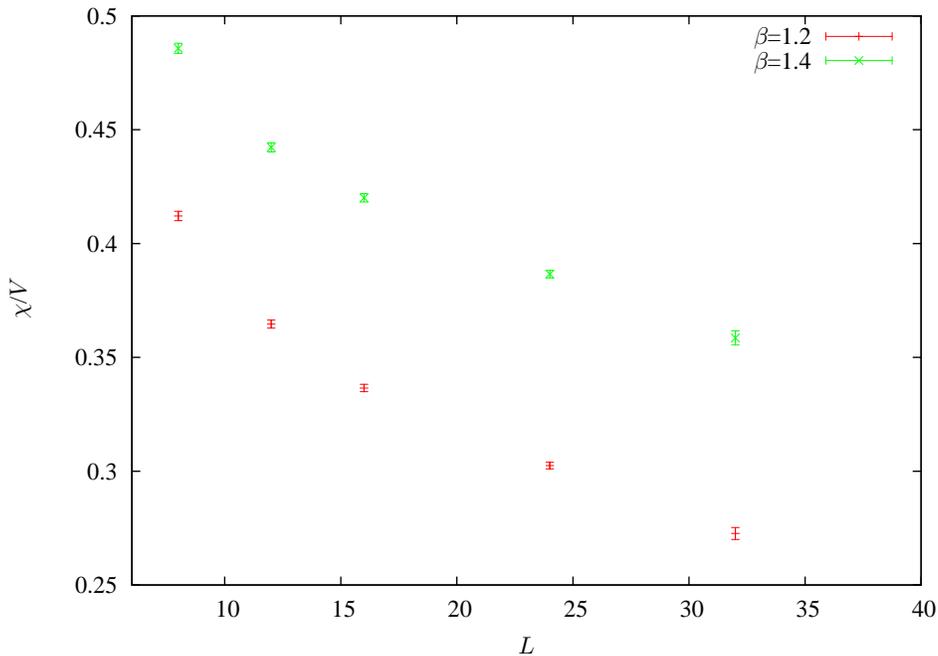}
\caption{$\chi/ V=\overline{\langle q^2 \rangle}$ versus the lattice size for $\beta=1.2$ and
  1.4. Notice that none of the  temperatures have reached the plateau asymptotic value.}
\label{fig:scaling_suscpetibility_b12_b14}
\end{figure}

The origin of this discrepancy with the standard theory can be understood
using Eq.~\ref{eq:scaling_e}. Notice from this equation that the scaling of
the zeros depends strongly on the behavior of the non-connected spin glass susceptibility,
so only with a divergence of this observable as the volume, we can  recover
$x_1=3$. However, for these two temperatures ($\beta=1.2$ and 1.4) this is not
the case (see Fig.  \ref{fig:scaling_suscpetibility_b12_b14}). Notice that
$\overline{\langle q^2 \rangle}=\chi/V$ has not reached the plateau asymptotic
value\footnote{In a spin glass phase, both the droplet as the RSB theory
  predict power law corrections on the lattice size, so the approach to the
  infinite volume values is really slow.}: Hence at these temperatures the
spin-glass susceptibility does not yet diverge as the volume.

\subsection{Behavior of the integrated density of zeroes}\label{sub:density}

\begin{figure}[!ht]
\centering
\includegraphics[width=.9\textwidth, angle=0]{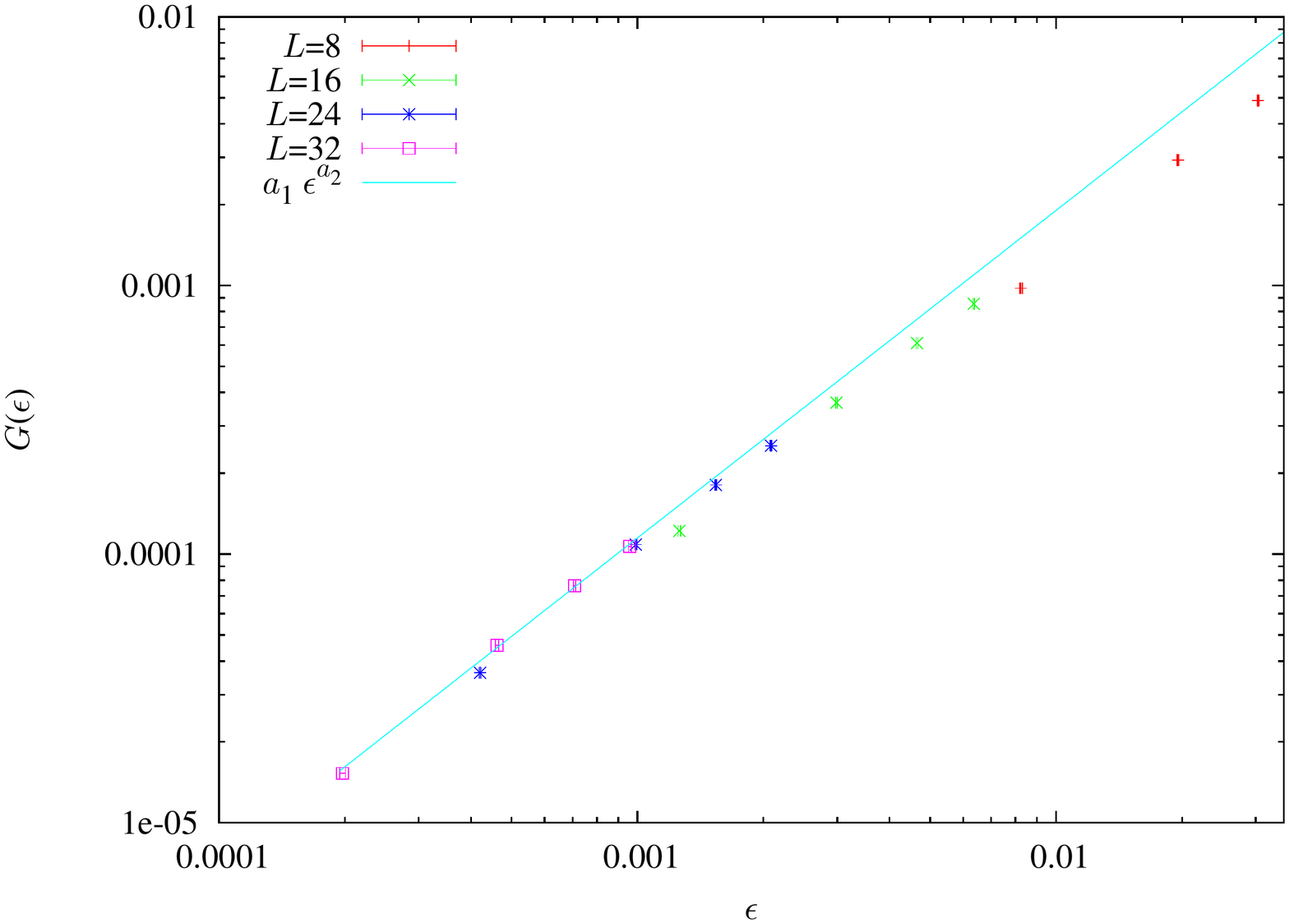}
\caption{Integrated density of zeros versus the zeros at the critical point. $a_2=1.16(2)$. }
\label{fig:density_bc}
\end{figure}

We will start our analysis of the integrated density of zeros by plotting this
density at the critical point in Fig.\ref{fig:density_bc}. One can see that
the largest lattices follow a pure power law as predicted by the theory. The
slope, on a log-log scale, of this straight line should correspond with an
exponent $a_2$. Fitting only the $L=32$ points we obtain $a_2=1.16(2)$ in good
agreement with the theory $a_2=1.116(2)$ (using Eq.~\ref{scaling_a2} and
$\eta=-0.375(10)$). To obtain this figure we have discarded in the fit the
first zero.\footnote{This phenomenon has been previously found in the
  literature. For example, the authors of \cite{janke:04} studied the
  anisotropic Ising model at the critical point and found a different behavior
  of the first zero in the study of the integrated density.  This model shows
  a spreading distribution of the zeros in the fugacity complex plane. The
  authors suggest that the effect of this spreading distribution of the zeros
  is modifying the behavior of the first zero. We have not computed the
  complete distribution of the complex zeros (only in the straight line $i
  \epsilon$), nevertheless, we know from reference~\cite{matsuda:08} that the
  zeros spread in the magnetic field complex plane, so it is quite natural to
  assume that we will have a similar (spreading) spatial distribution of the
  zeros in $\epsilon$.  Another possible explanation is that the behavior of
  the integrated density of zeros as $j/L^3$ is only asymptotic.  These
  anomalies affect only the lower order zeros. Notice that this phenomenon
  affects only to the pre-factor of the power law of the smallest zeros. We have
  seen in subsection \ref{sub:scaling_tc} that the first zero scales with the
  right power law. We thank R. Kenna for interesting comments regarding this
  behavior.}

For large $L$ we should expect a
good collapse of all points in the same power law curve: the non collapsing
part (small $L$ in the figure) is due to the presence of scaling corrections
(which we also found in the previous sections).

\begin{figure}[!ht]
\centering
\includegraphics[width=.9\textwidth, angle=0]{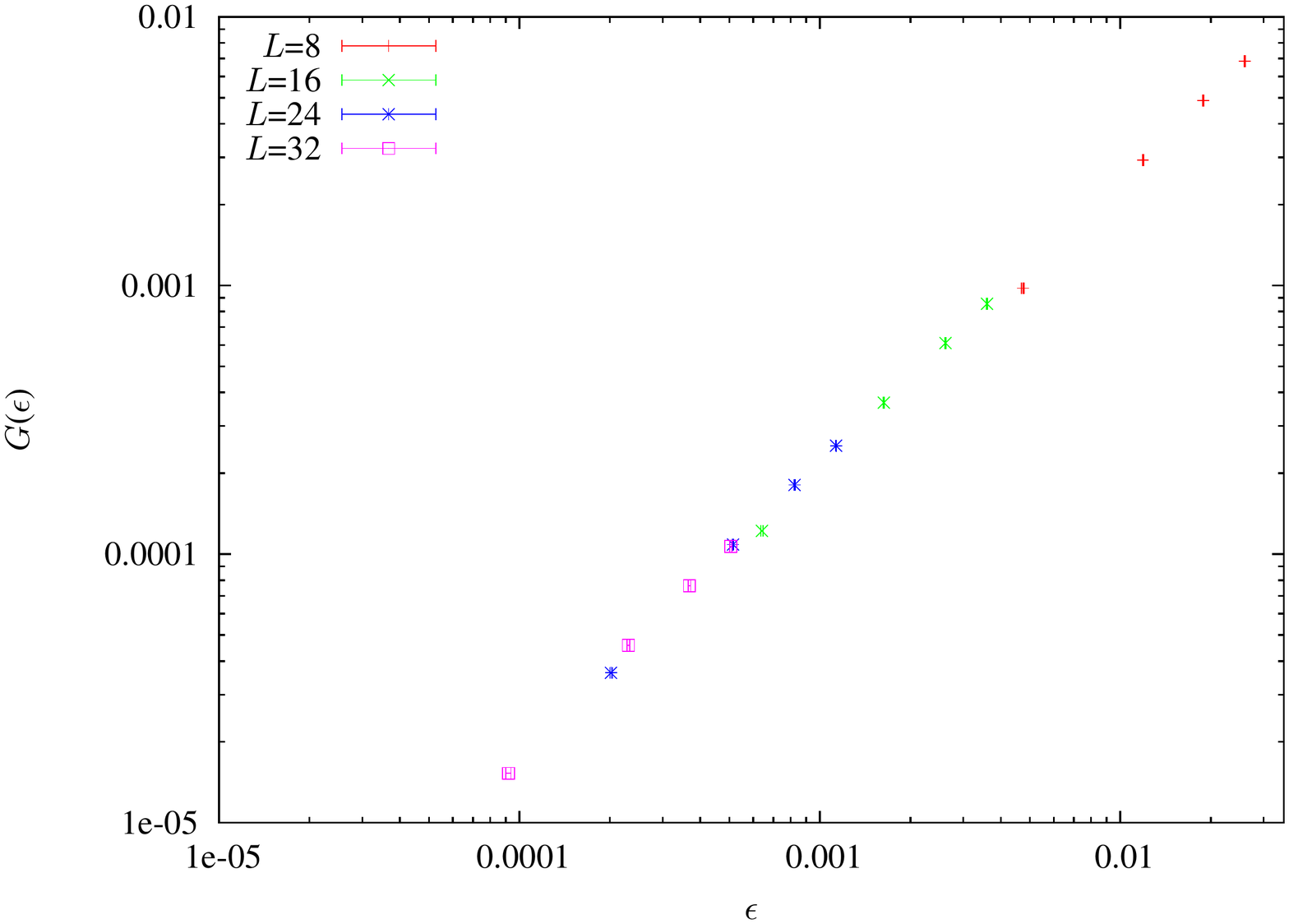}
\caption{Integrated density of zeros versus the zeros for $\beta=1.2$.}
\label{fig:density_b12}
\end{figure}

Now we will check the theoretical predictions for the integrated density of
zeros in the broken phase, which predict a linear behavior in the perturbing
parameter $\epsilon$. Notice that in our case the margins between the critical
point and the broken phase are tight since in the infinite volume limit we
will see a behavior $\epsilon^{1.116}$ at the critical point which changes just
below $T_c$ to $\epsilon$ (of course, this is due to the value of the $\eta$ exponent).

\begin{figure}[!ht]
\centering
\includegraphics[width=.9\textwidth, angle=0]{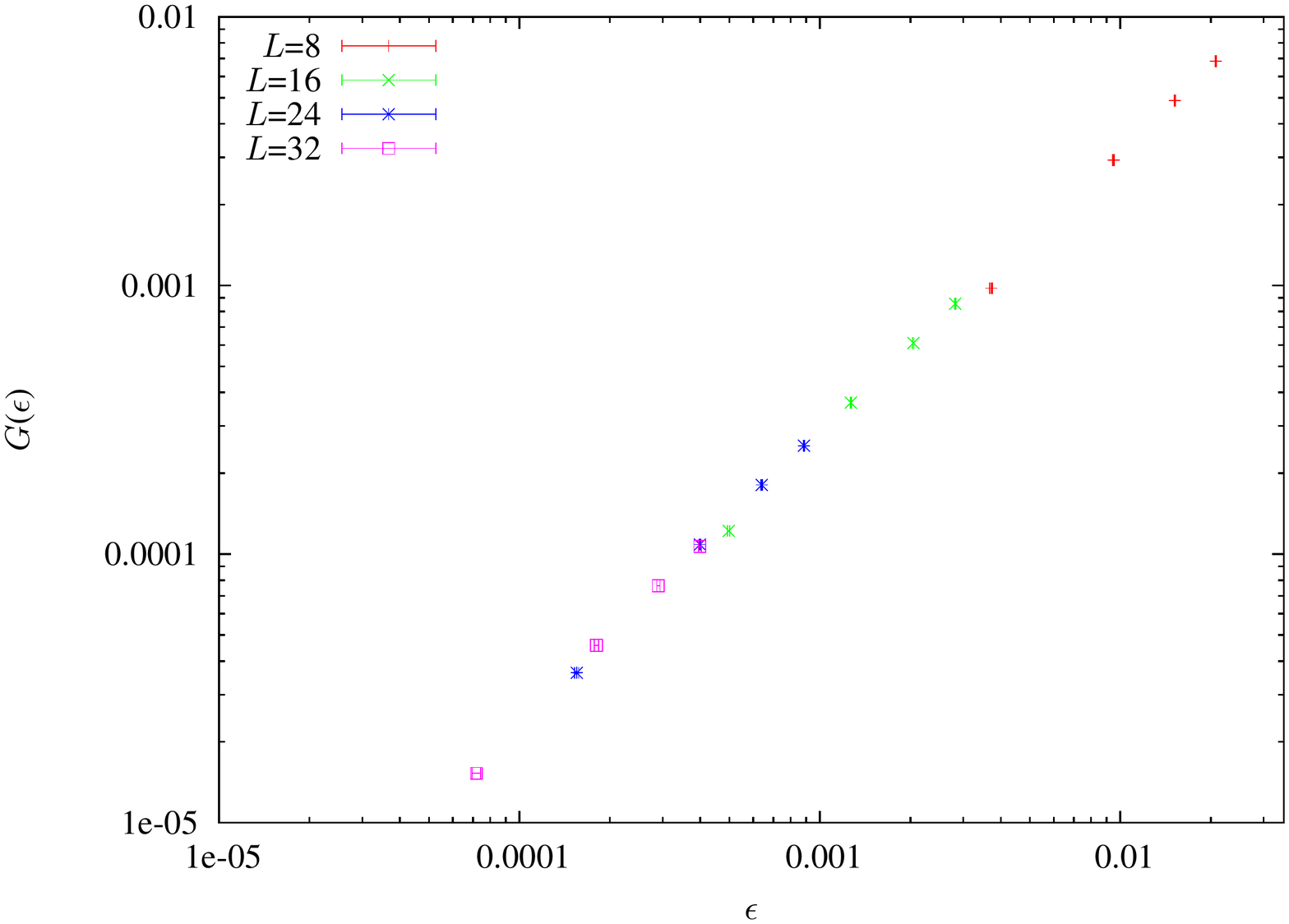}
\caption{Integrated density of zeros versus the zeros at $\beta=1.4$.}
\label{fig:density_b14}
\end{figure}

In Figs \ref{fig:density_b12} and
\ref{fig:density_b14} we show that the data nearly follow a linear behavior
of the integrated density deep in the spin glass phase (more concretely at
$\beta=1.2$ and $\beta=1.4$), in particular for
$L\ge 24$. The non-collapsing part of the curve is produced by the
presence of scaling corrections as at the critical point.

However, it is easy to show that if the zeros do not follow (for the lattice
sizes simulated), in the broken phase, a scaling as the inverse of the volume,
then the integrated density of zeros does not follow exactly a linear
behavior, since $a_2=D/x_1$. We have discussed at the end of
Sec.~\ref{sec:scaling_low} that this lack of $1/V$ behavior is related to a
susceptibility that is not yet diverging as the volume.

In sec.~\ref{sec:scaling_low} we have found an exponent $x_1=2.842(11)$ for
$\beta=1.2$ and $x_1=2.844(10)$ for $\beta=1.4$, which implies that
$a_2=1.056(4)$ and $a_2=1.055(3)$ for $\beta=1.2$ and $\beta=1.4$ respectively. 

\begin{figure}[!ht]
\centering
\includegraphics[width=.9\textwidth, angle=0]{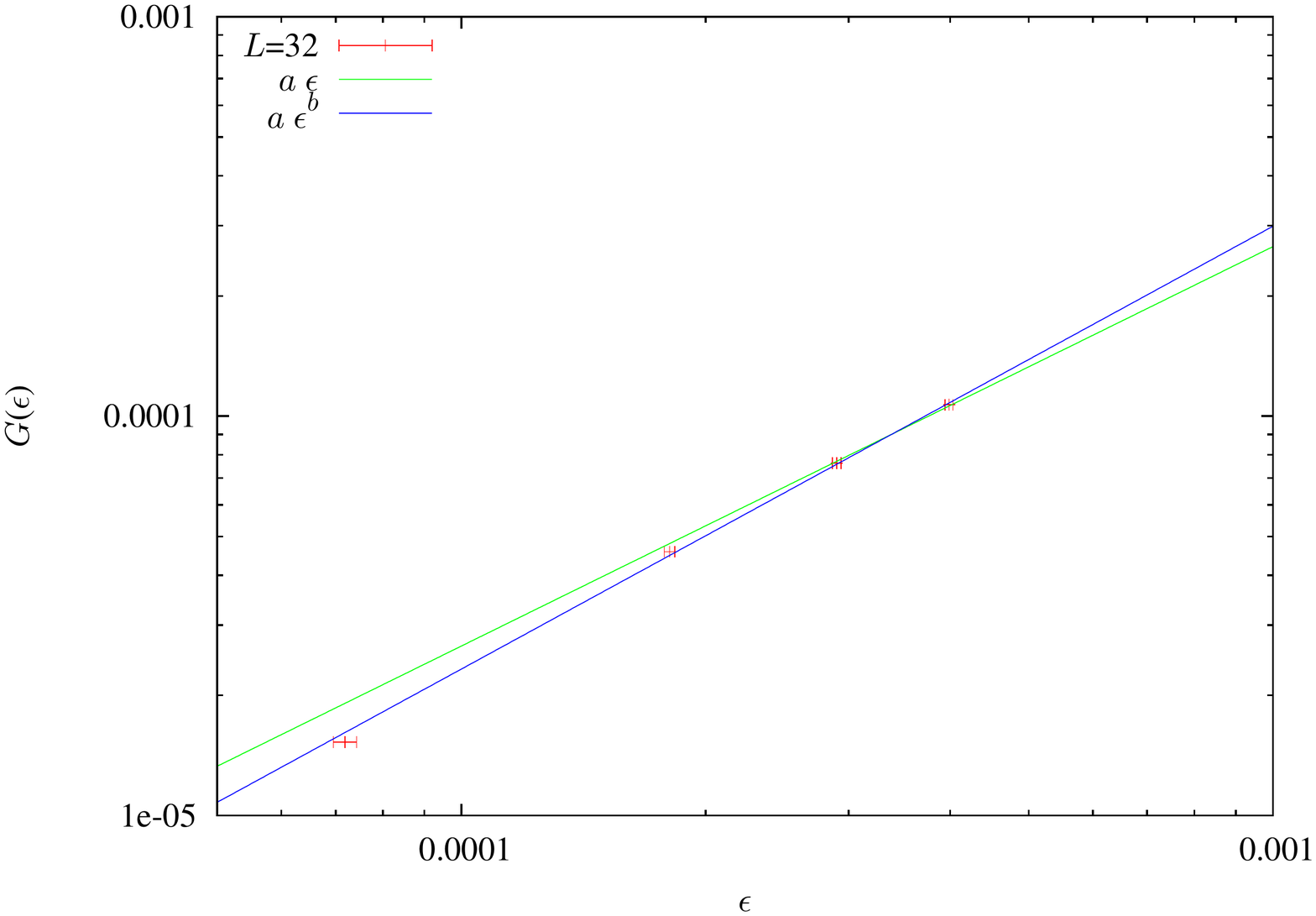}
\caption{Integrated density of the zeros, for the largest lattice $L=32$ and
  the lowest temperature $\beta=1.4$. Notice that we are almost, but not in,
  the linear regime. The data are well fitted with $b=1.068(10)$.}
\label{fig:density_b14_firstzero_L32}
\end{figure}

In Fig. \ref{fig:density_b14_firstzero_L32} we show the behavior of the
integrated density of zeros computed for our largest lattice ($L=32$) and
lowest temperature, ($\beta=1.4$). Notice the points are not lying on a straight line. 
A fit to $a_1
\epsilon^{a_2}$ works well, with $a_2$ consistent with the value computed from
$x_1$ ($a_2=1.068(10)$).  So we have obtained, numerically, $a_2=1.16(2)$, at
the critical point which has changed to $1.068(10)$ in the broken
phase.\footnote{We can do the same analysis with the $x_1$ exponent: we have
  obtained at the critical point $x_1= 2.67(7)$, which should change in the
  broken phase to $x_1=3$, although we actually see with our numerical data
  $x_1=2.842(11)$.}
\begin{figure}[htb]
\centering
\includegraphics[width=.9\textwidth, angle=0]{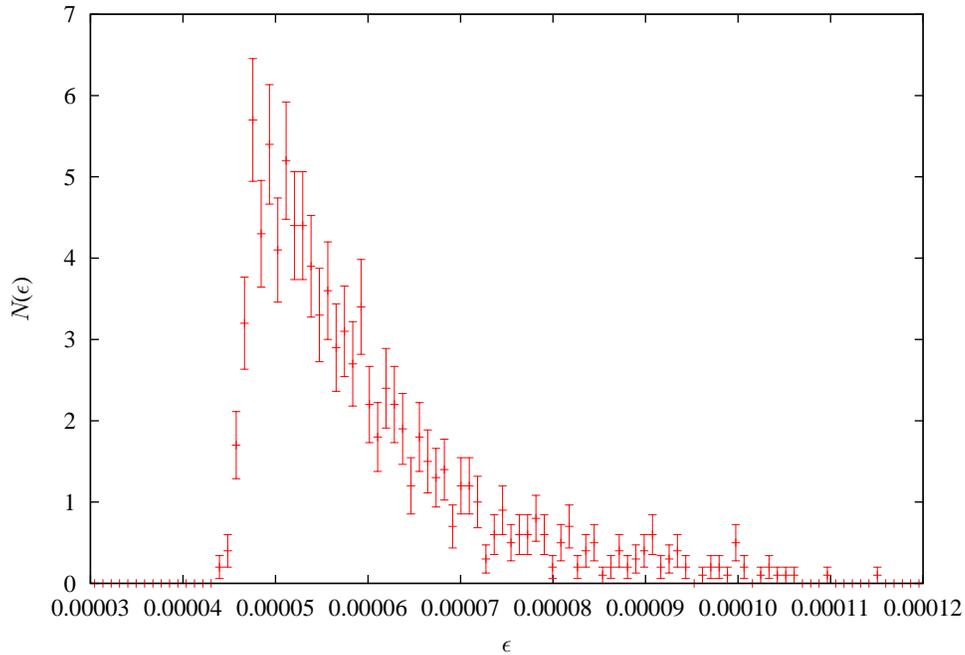}
\caption{Histogram ($N(\epsilon)$ versus $\epsilon$) for the 1000 first zeros
  computed for $L=32$ and $\beta=1.4$. Notice the lack of symmetry of the
  histogram and the presence of events for large values of the zeros.}
\label{fig:hist_L32_b14}
\end{figure}

\begin{figure}[htb]
\centering
\includegraphics[width=.9\textwidth, angle=0]{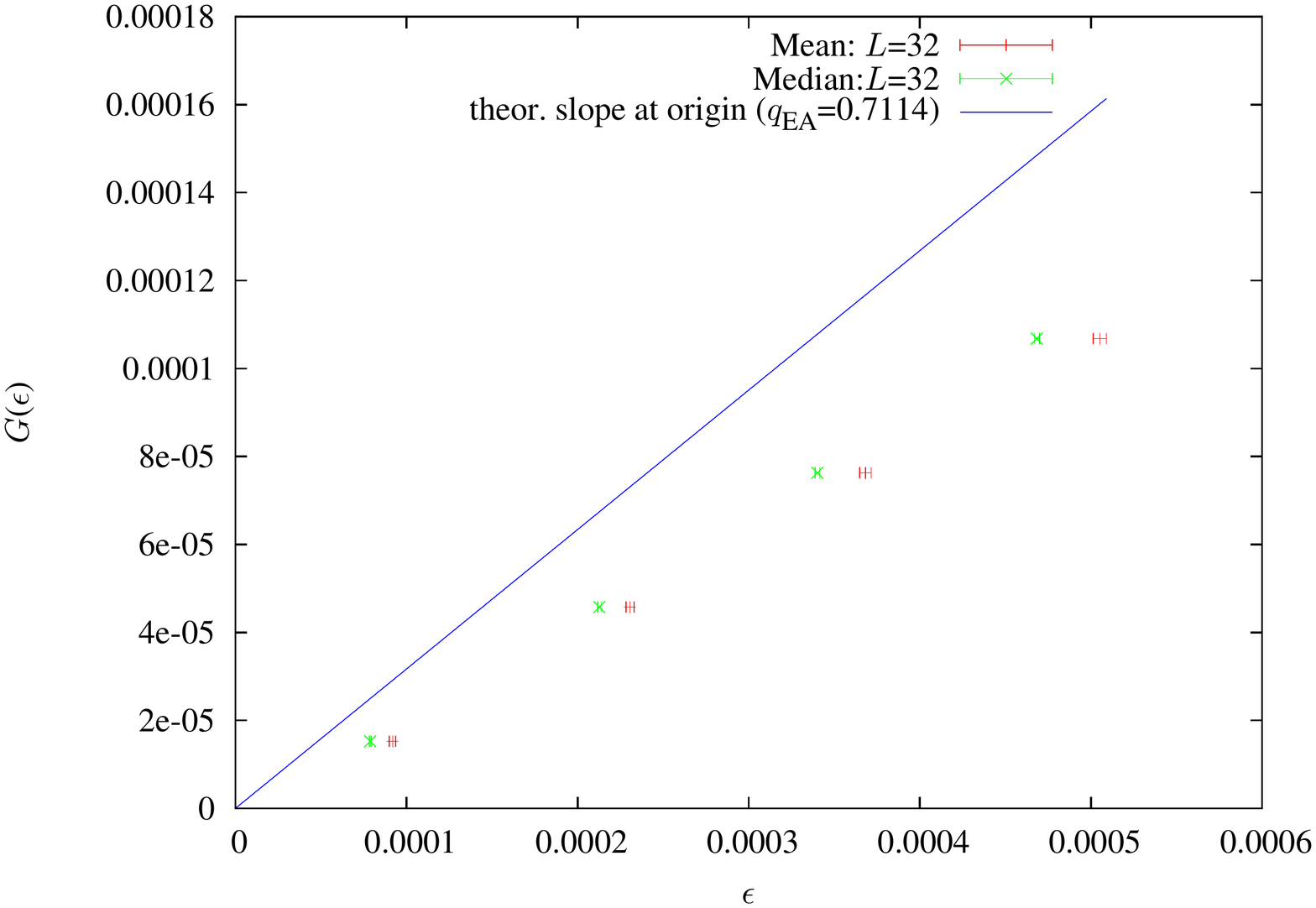}
\caption{Integrated density of the zeros, for the largest lattice $L=32$ and
  temperature $\beta=1.2$ using the average of zeros. We have also plotted the median values.
We have marked the expected slope at the origin,
  using the Edwards-Anderson order parameter computed in Ref.~\cite{janus:10}
  for the $L=32$ lattice.}
\label{fig:density_b12_firstzero_L32_comp}
\end{figure}

\begin{figure}[htb]
\centering
\includegraphics[width=.9\textwidth, angle=0]{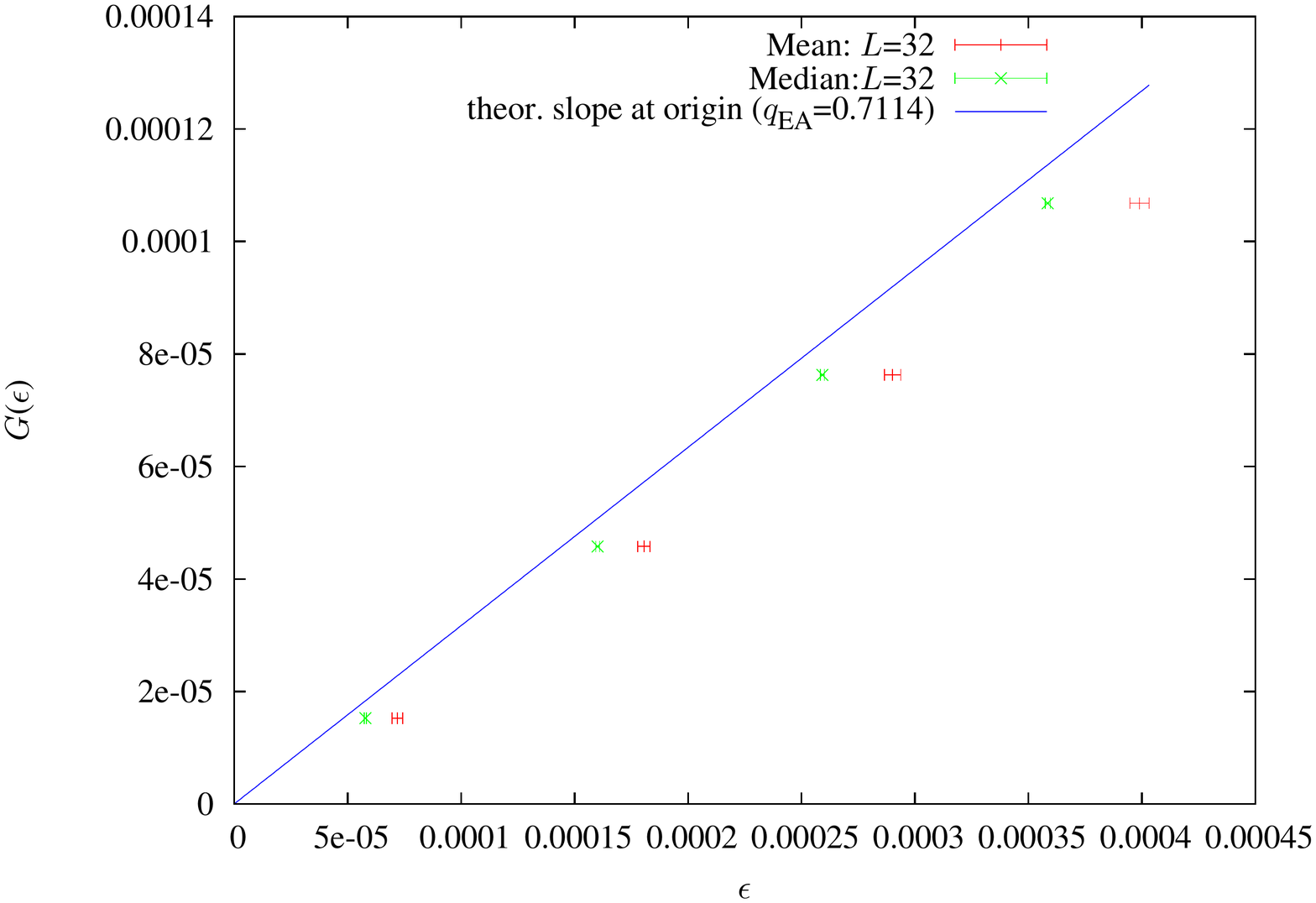}
\caption{Integrated density of the zeros, for the largest lattice $L=32$ and
  lowest temperature $\beta=1.4$ using the average of zeros. We have also
  plotted the median values. We have marked the expected slope at the origin,
  using the Edwards-Anderson order parameter computed in Ref.~\cite{janus:10}
  for the $L=32$ lattice.}
\label{fig:density_b14_firstzero_L32_comp}
\end{figure}

In this situation, we cannot compute the order parameter directly from the
linear behavior of the integrated density since we are not observing a fully
linear behavior. Hence, we confront our numerical data for $G(\epsilon)$
against the theoretical prediction for really small $\epsilon$, which is
$G(\epsilon)= (\beta q_\mathrm{EA}/\pi) \epsilon$. It is interesting to note
(see Refs.~\cite{janke:01} and \cite{gordillo:12}) that we can recover the exact
slope for a given lattice size if we substitute the value of the order
parameter computed for this lattice size. We have followed this advice, and we
show in Figs. \ref{fig:density_b12_firstzero_L32_comp} and
\ref{fig:density_b14_firstzero_L32_comp} our data for $G(\epsilon)$ at
$\beta=1.2$ and 1.4 showing $L=32$ data. In addition we have plotted the
asymptotic slope using the order parameter ($q_\mathrm{EA}$) computed for
$L=32$ lattices for these two temperatures in Ref.~\cite{janus:10}. Notice
that we have a slow approach to the right slope, but also that the overall
picture seems to be correct.

In order to gain a better understanding of this effect, we have computed the
density of zeros not with the average of the sample zeros but with the median
of the probability distribution of the zeros.\footnote{The probability
  distributions one usually finds in disordered systems present long tails due
  the presence of rare events, hence, the study of the median of this kind of
  distributions is also very useful (see for example~\cite{fernandez:08, billoire:11}).}

We show in Fig.~\ref{fig:hist_L32_b14} the histogram of the 1000 first zeros
computed on the $L=32$ lattice at $\beta=1.4$.  Notice from this figure the
asymmetry of the histogram and the presence of events at large values of the
zeros (which induces a large and strongly fluctuating value of the mean).

One can see in Figs. \ref{fig:density_b12_firstzero_L32_comp} and
\ref{fig:density_b14_firstzero_L32_comp} that the median data produce  an
improved scaling, compared  with those obtained from the mean, when comparing
the data with the analytical prediction (slope provided by $q_\mathrm{EA}$).

For the sake of completeness, we can cite that the integrated density of zeros
using the medians does not behave completely linearly but with a law
$\epsilon^{1.06(2)}$ (for $\beta=1.4$).

\section{Conclusions}\label{CONCLUSIONS}

By studying the complex singularities linked with the overlap we have obtained
a clear picture of the critical region and of the low temperature phase fully
compatible with that obtained by other more standard  approaches.

In particular, we have studied the behavior of the individual zeros as well as
the integrated density at the critical point. In both cases we have obtained
good values for the $\eta$ exponent and we have seen that the data are
compatible with the corrections to scaling published in the
literature~\cite{hasenbusch:08b}.

Finally, we have checked the scaling laws in the spin-glass phase, obtaining
strong scaling corrections as found previously~\cite{janus:10}. In addition we
have obtained, by monitoring the behavior of the integrated density, a
compatible picture using the zeros with that obtained from the order parameter
of the model ($q_\mathrm{EA}$) computed in finite volumes with
standard methods. We have also shown that the use of the median instead of the mean 
improves the overall picture.

\section*{Acknowledgments}

The authors wish to thank the Janus Collaboration for allowing us to analyze
their data.  We also wish to thank R. Kenna for interesting comments on the
manuscript.  J.J.R.L. acknowledges support from Research Contracts No.
FIS2007-60977 (MICINN), GR10158 (Junta de Extremadura) and
PIRSES-GA-2011-295302 (European Union).  D.Y. acknowledges support from the
European Research Council under the European Union's Seventh Framework
Programme (FP7/2007-2013, ERC grant agreement no. 247328).  R.A.B., J.M.-G.,
J.M.G.-N. and D.Y. acknowledge support from MICINN, Spain (contract no.
FIS2009-12648-C03). R.A.B. and J.M.G. acknowledge support from the Formaci\'on de Personal
Investigador (FPI) program (Diputaci\'on General de Arag\'on, Spain).

\section*{References}
\providecommand{\newblock}{}

\end{document}